\documentclass[%
 reprint,
 superscriptaddress,
 showkeys,
nofootinbib,
 amsmath,amssymb,
 aps,
 prd,
]{revtex4-2}

\usepackage[utf8]{inputenc}
\DeclareUnicodeCharacter{2500}{\textemdash{}}

\usepackage[normalem]{ulem}
\usepackage{xcolor}
\usepackage{cancel}

\usepackage{textcomp}
\usepackage{soul}
\usepackage{float}
\usepackage[caption=false]{subfig} 
\usepackage{gensymb}
\usepackage{graphicx}
\usepackage{dcolumn}
\usepackage{bm}
\usepackage[hypertexnames=true,colorlinks=true,urlcolor=blue,bookmarks=true,citecolor=blue,breaklinks=true]{hyperref}
\usepackage[dvipsnames]{xcolor}
\usepackage{booktabs}
\usepackage{makecell}
\usepackage{array}
\usepackage{multirow}
\usepackage[normalem]{ulem}
\newcolumntype{C}[1]{>{\centering\arraybackslash}p{#1}}

\begin{document}

\preprint{APS/123-QED}

\title{Equation of State Classification of Core-Collapse Supernovae with a Network of Gravitational-Wave Detectors}

\author{A.~Akhmetali}
\email{almat.akhmetali@nu.edu.kz} 
\affiliation{Energetic Cosmos Laboratory, Nazarbayev University, 010000 Astana, Kazakhstan} 
\affiliation{Department of Physics, Nazarbayev University, 010000 Astana, Kazakhstan}

\author{Y.~S.~Abylkairov}
\email{sultan.abylkairov@nu.edu.kz}
\affiliation{Energetic Cosmos Laboratory, Nazarbayev University, 010000 Astana, Kazakhstan} 
\affiliation{School of Artificial Intelligence and Data Science, Astana IT University, 010000 Astana, Kazakhstan}

\author{J.~A.~Font}
\affiliation{Departamento de Astronomía y Astrofísica, Universitat de València, Avinguda Vicent Andrés Estellés 19, 46100 Burjassot (Valencia), Spain}
\affiliation{Observatori Astronòmic, Universitat de València, Catedrático José Beltrán 2, 46980, Paterna, Spain}

\author{M.~Zanolin}
\affiliation{Department of Physics and Astronomy, Embry-Riddle Aeronautical University, Prescott, AZ 86301, USA}

\author{E.~Abdikamalov}
\affiliation{Energetic Cosmos Laboratory, Nazarbayev University, 010000 Astana, Kazakhstan}
\affiliation{Department of Physics, Nazarbayev University, 010000 Astana, Kazakhstan}

\date{\today}

\begin{abstract}
Gravitational waves from core-collapse supernovae may provide a unique probe of the equation of state (EOS) of dense nuclear matter. In previous work, we assessed the potential of machine-learning methods for EOS classification for rapidly rotating supernovae using individual gravitational-wave detectors. Here, we extend the framework to detector networks and investigate the impact of detector configuration, source distance, arrival time, and sky location on the classification. Using convolutional neural networks, we perform 10-class EOS classification directly from noisy detector data for both second- and third-generation observatories. We find that detector geometry strongly influences classification performance. For Galactic events, current detector networks can achieve promising classification performance, but the reach depends sensitively on the detector network antenna response. Adding detectors improves time-averaged performance and produces more uniform classification across the sky. Third-generation observatories extend the classification reach to 100 kpc for favorable source orientations but do not remove the dependence on detector geometry. 
\end{abstract}


\maketitle

\section{Introduction}
\label{sec:intro}

Gravitational waves (GWs) have emerged as a powerful probe of the most extreme astrophysical phenomena. Observations of compact binary mergers are now routine, advancing our understanding of neutron stars and black holes \citep{Abac26GWTC5}. Core-collapse supernovae (CCSNe) are another major class of expected GW sources \citep{mezzacappa25colloquium}. A detection of GWs from a nearby CCSN would offer an unobstructed view of the supernova central engine, revealing information inaccessible to electromagnetic observations alone \citep{Mueller26GWREview}. Current GW observatories are expected to detect CCSN signals from Galactic events \citep{abbott2020optically, szczepanczyk2023optically, Szczepanczyk21Detecting}, while future detectors will extend the detection horizon and improve source characterization \citep{srivastava19detection, Srivastava22, Akhmetali26PE}.

CCSNe occur when massive stars exhaust their nuclear fuel and their cores become gravitationally unstable. Collapse proceeds until nuclear densities are reached, at which point the inner core rebounds and launches a shock wave. Because this shock rapidly loses energy through nuclear dissociation and neutrino emission, it stalls shortly after formation and must be revived for a successful explosion to occur \citep{Janka16Physics, muller20hydrodynamics, Burrows24Physical}. In the standard neutrino-driven paradigm, a fraction of the neutrino luminosity emitted by the proto-neutron star (PNS) is absorbed behind the shock and contributes to its revival \citep{Janka01Conditions, burrows13colloquium}. Additional support may arise from hydrodynamic instabilities, including convection and the standing accretion shock instability (SASI), as well as from perturbations generated in the progenitor prior to collapse \citep{blondin03stability, foglizzo06neutrino, Couch15Three, Mueller17Supernova, Kazeroni20impact, Telman24Convective}. In rapidly rotating progenitors, rotational energy can instead power magnetorotational explosions through magnetic stresses and bipolar outflows \citep{burrows:07b, moesta:14b, obergaulinger:20, kuroda:20}.

Because GWs are generated by multidimensional motions in the stellar core, they directly trace the dynamics of the explosion engine \citep{abdikamalov22gravitational}. In slowly rotating CCSNe, the dominant GW emission arises from PNS oscillations excited by convection and accretion flows, together with contributions from SASI activity, anisotropic neutrino emission, and explosion asymmetries \citep{Murphy09Model, Ott13General, Yakunin15GW, TorresForne19Universal, Mezzacappa23Core, Vartanyan23Gravitational, Sotani24Universality, Choi24GW, Birnholtz13GW_jet, Soker23GWJJ,Tseneklidou26}. In rapidly rotating progenitors, the GW signal is qualitatively different \citep{Powell26Gravitational}. Centrifugal deformation of the collapsing inner core produces a strong burst of GW emission at core bounce, followed by a ring-down phase associated with quadrupolar oscillations of the newly formed PNS \citep{Fuller15SNseismology}. Rapid rotation may subsequently trigger non-axisymmetric instabilities that generate longer-lived GW emission \citep{Scheidegger08, Shibagaki20new, Pan21Stellar}, while additional contributions can arise from resonant interactions involving PNS oscillation modes \citep{Cusinato:26}. Because the GW signal is shaped by the physical conditions in the collapsing core, it provides a unique probe of the supernova central engine and offers the possibility of inferring key source properties from observations \citep{Logue12Inferring, abdikamalov:14, pajkos19, pajkos21, Powell24Determining, pastor24, nunes2024deep, Villegas25Parameter, Avila26}.

While rapidly rotating CCSNe are expected to be rare, possibly comprising only $\sim 1 \%$ of all events \citep{Heger05Presupernova, woosley:06, popov:12}, the rotating bounce and early post-bounce signal offers an important advantage: it can be modeled with axisymmetric simulations at comparatively modest computational cost \citep{ott:07cqg, Dimmelmeier08, ott12correlated}. This enables the construction of large waveform catalogs spanning a wide range of progenitors, rotation rates, and equation-of-state (EOS) models. Such datasets are particularly well suited for machine-learning (ML) approaches, which require large numbers of training examples. Recent studies have demonstrated that ML techniques can extract source properties directly from CCSN GW signals \citep{mitra23, mitra24, Abylkairov24Evaluating, Sakan25Probing}. In particular, these methods can constrain the rotation of the collapsing core \citep{Akhmetali26PE, Akhmetali26PE_DN} and probe the EOS of high-density matter \citep{richers:17, edwards17, chao22determining, Cassallas-Lagos26, Llorens-Monteagudo26}.

In a recent study \citep{Akhmetali26Toward}, we investigated EOS inference from the GW signals generated during rotating core collapse and bounce. Building upon earlier work \citep{mitra24, Abylkairov24Evaluating, abylkairov2025assessing}, we increased the realism of the analysis by incorporating real detector noise from the LIGO O4a observing run, accounting for uncertainties in the bounce time, and extending the waveform catalog to multiple progenitor models spanning a broad range of stellar masses. Using ML techniques, we demonstrated that EOS classification remains feasible despite these additional sources of uncertainty. However, the study reported in~\citep{Akhmetali26Toward} considered only individual GW detectors and therefore neglected the additional information available from coordinated observations by detector networks.

In practice, gravitational-wave (GW) observations are conducted using networks of detectors distributed across different geographic locations. The sensitivity, sky coverage, polarization reconstruction capability, and duty cycle of such a network depend critically on the relative positions and orientations of its constituent instruments. The current global gravitational-wave detector network includes the two Advanced LIGO detectors \citep{Aasi:2014jea}, Advanced Virgo \citep{Acernese15}, and KAGRA \citep{Aso13}, with LIGO India expected to join in the near future \citep{Saleem22}. Future third-generation facilities, including the Einstein Telescope (ET) \citep{2026:ET} and Cosmic Explorer (CE) \citep{Reitze19, Evans21}, will provide a significant increase in sensitivity and correspondingly enlarge the volume accessible to CCSN detections \citep{srivastava19detection, Srivastava22}. Recent studies have demonstrated that detector networks can substantially improve the recovery of source properties from CCSN GW signals by combining information from multiple detectors and reconstructing both GW polarizations \citep{Bruel23Inference, Akhmetali26PE_DN}. 

Whether similar gains can be achieved for EOS inference from the bounce and early ring-down signals of rapidly rotating CCSNe remains an open question. Signal amplitude depend strongly on source orientation and detector response. Combining information from multiple detectors may therefore improve EOS classification by reducing orientation-dependent degeneracies and increasing the effective signal-to-noise ratio (SNR). At the same time, the extent of these improvements depends on detector sensitivity, detector network geometry, and source distance.

In this work, we extend our previous EOS-classification study \citep{Akhmetali26Toward} to GW detector networks. We investigate how detector network configuration influences the ability to infer the nuclear EOS from the bounce and early post-bounce GW signal of rapidly rotating CCSNe. We consider both current and future GW observatories, including detector networks composed of Advanced LIGO Plus (A+), ET, and CE detectors. As in our previous work, we employ ML methods trained on a large catalog of rotating-collapse waveforms spanning multiple progenitors, rotational configurations, and EOS models. We incorporate simulated detector noise, uncertainties in the bounce time, and variations in source orientation. Our goal is to quantify the extent to which detector networks improve EOS inference, determine the distance reach achievable by different detector network configurations, and assess the prospects for constraining dense-matter physics with future GW observations of rapidly rotating CCSNe.

The remainder of this paper is organized as follows. In Section~\ref{sec:detectors}, we describe the GW detector networks considered in this study. Section~\ref{sec:methods} presents the waveform dataset and the inference methodology. Our results are presented and discussed in Section~\ref{sec:results}. Finally, Section~\ref{sec:conclusion} summarizes the main findings and offers concluding remarks. Technical details on the methodology employed in this work are provided in Appendix~\ref{sec:appendix}. 

\section{Detector Network}
\label{sec:detectors}

Ground-based GW observatories are commonly divided into second-generation (2G) and third-generation (3G) detectors. The current 2G detector network comprises the two Advanced LIGO interferometers located in Hanford and Livingston, USA~\citep{Aasi:2014jea}, Advanced Virgo in Italy~\citep{Acernese15}, and KAGRA in Japan~\citep{Aso13}. The LIGO interferometers have 4-km arms, whereas Virgo and KAGRA have 3-km arms. The global detector network is expected to expand further with the addition of LIGO Aundha in India~\citep{Saleem22}, a 4-km LIGO-like interferometer currently under construction and anticipated to begin operations around 2030. These observatories operate as a coordinated detector network, carrying out joint observing runs interspersed with periods of commissioning and detector upgrades. This progressive development is intended to improve the sensitivity of the individual instruments toward their respective design goals~\citep{Abbot20}.

Third-generation observatories are designed to extend the sensitivity and astrophysical reach of ground-based GW astronomy well beyond that of the current generation~\citep{ET_CE_noise}. The two principal 3G detector concepts are the Einstein Telescope in Europe and Cosmic Explorer in the United States. Two configurations are currently under consideration for the ET: a two-site L-shaped configuration (2L), consisting of two 10-km interferometers separated by approximately 1200 km, and a triangular configuration comprising three interferometers arranged underground~\citep{ET:comparison}. In this work, we adopt the triangular ET-D configuration. It consists of six 10-km arms forming an equilateral triangle, with opening angles of $60^\circ$. The triangular geometry provides three interferometric data streams and enables multiple measurements of the GW signal within a single facility. The ET-D design also employs a xylophone configuration, in which dedicated low- and high-frequency interferometers use different technologies to reduce seismic, thermal, and quantum noise across the broad frequency range relevant to GW observations~\citep{2026:ET}.

The CE concept, in contrast, is based on conventional L-shaped interferometers with substantially longer arms. The proposed facilities include 20-km and 40-km arm-length configurations, with the latter representing the ultimate CE design. The increased arm length, combined with improved optical configurations and noise-reduction technologies, is expected to provide a substantial improvement in strain sensitivity relative to 2G observatories~\citep{ET_CE_noise}. Together, ET and CE will form the basis of the next generation of ground-based GW observations, providing increased sensitivity, broader frequency coverage, and improved source localization capabilities.

\begin{figure}[t!]
\centering
\includegraphics[width=1\linewidth]{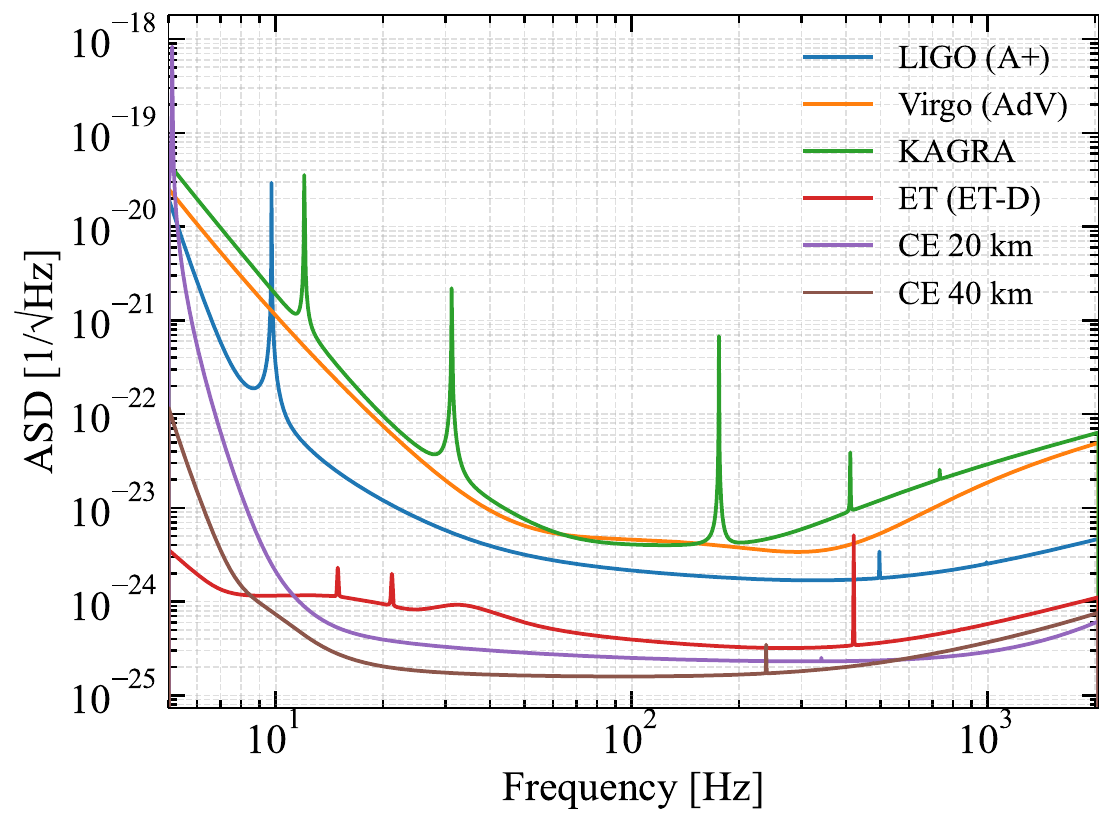}
\caption{ASDs as a function of frequency for the ground-based GW detectors considered in this study.}
\label{fig:ASDs}
\end{figure}

Within an Earth-fixed reference frame, the geometry of an interferometric detector is specified by the location of its beam splitter and the orientations of its arms. The corresponding geographical and geometrical parameters adopted for the 2G and 3G observatories considered in this work are summarized in Table~\ref{tab:detectors}. Since the locations and orientations of the future 3G observatories have not yet been finalized, we adopt representative configurations based on those used in previous studies. Following~\citet{Bruel23Inference}, we place the ET at the Virgo site and choose its overall orientation arbitrarily. For the CE detector network, we use the proposed sites in Idaho and New Mexico, together with the corresponding orientations, following~\citet{Borhanian21}. These configurations are used as representative realizations of future 3G detector networks and are not intended to reflect their eventual final designs.

\begin{table*}[t]
    \centering
    \resizebox{\textwidth}{!}{
    \begin{tabular}{c|c|c|c|c|c|c|c|c|c|c|c|c}
        \hline
        Detector & $\lambda$ [$^\circ$] & $\phi$ [$^\circ$] & $h$ [m] & $\Psi_1$ [$^\circ$] & $\Psi_2$ [$^\circ$] &
        Detector & $\lambda$ [$^\circ$] & $\phi$ [$^\circ$] & $h$ [m] & $\Psi_1$ [$^\circ$] & $\Psi_2$ [$^\circ$] & $\Psi_{\rm GC}$ [$^\circ$] \\
        \hline
        LIGO Hanford (H)     & -119.41 & 46.46 & 142.55 & 324.00 & 234.00 &
        Einstein Telescope 1 (ET) & 10.50 & 43.63 & 0.0 & 89.95 & 29.96 & -- \\
        
        LIGO Livingston (L)  & -90.77 & 30.56 & -6.57 & 252.28 & 162.28 &
        Einstein Telescope 2 (ET) & 10.63 & 43.63 & 0.0 & 330.04 & 270.05 & -- \\
        
        Virgo (V)            & 10.50 & 43.63 & 51.88 & 19.43 & 289.43 &
        Einstein Telescope 3 (ET) & 10.57 & 43.71 & 0.0 & 210.00 & 150.00 & -- \\
        
        KAGRA (K)            & 137.31 & 36.41 & 414.18 & 60.40 & 330.40 &
        Cosmic Explorer 40 km (CE40) & -112.83 & 43.83 & 0.0 & 180.00 & 90.00 & 27.02 \\
        
        LIGO Aundha (A)      & 77.03 & 19.61 & 0.0 & 332.38 & 242.38 &
        Cosmic Explorer 20 km (CE20) & -106.48 & 33.16 & 0.0 & 240.00 & 150.00 & 83.04 \\
        \hline
    \end{tabular}
    }
    \caption{Geographical and geometrical parameters of the ground-based observatories considered in this work. The left and right halves of the table correspond to the 2G and 3G detectors, respectively. The longitude $\lambda$ and latitude $\phi$ specify the beam-splitter (corner-station) location, while $h$ is the elevation above the WGS84 reference ellipsoid. The angles $\Psi_1$ and $\Psi_2$ give the orientations of the interferometer arms measured clockwise from local North. For the CE detectors, $\Psi_{\rm GC}$ specifies the orientation of the first interferometer arm relative to the great circle connecting the two CE sites. The parameters for LIGO Hanford, LIGO Livingston, Virgo, KAGRA, and LIGO Aundha are adopted from~\citep{lalsuite}. Because the locations and orientations of the 3G observatories have not yet been finalized, we adopt representative configurations. Following~\citep{Bruel23Inference}, the ET is placed at the Virgo site with an arbitrary overall orientation. The CE sites and orientations follow~\citep{Borhanian21}, with the 40-km detector located in Idaho and the 20-km detector in New Mexico.}
    \label{tab:detectors}
\end{table*}

In addition to their geometrical response, the sensitivity of the detectors is determined by their noise spectra. For the 2G detector network, we use the projected design sensitivity of Advanced LIGO (A+)~\citep{barsotti2018a+}\footnote{\texttt{aLIGOAPlusDesignSensitivityT1800042} from \texttt{PyCBC}}, the Advanced Virgo design sensitivity~\citep{Abbot20}\footnote{\texttt{AdvVirgo} from \texttt{PyCBC}}, and the KAGRA design sensitivity~\citep{Abbot20}\footnote{\texttt{KAGRADesignSensitivityT1600593} from \texttt{PyCBC}}. For the 3G detector network, we adopt the ET-D sensitivity curve~\citep{Hild2011ET}\footnote{\texttt{EinsteinTelescopeP1600143} from \texttt{PyCBC}} for ET and compact-binary optimized CE sensitivity curves~\citep{Srivastava22}\footnote{We use \texttt{CosmicExplorerP1600143} for the 40-km CE detector and \texttt{CosmicExplorerPessimisticP1600143} for the 20-km detector, both implemented in \texttt{PyCBC}.} The resulting amplitude spectral densities (ASDs) are shown in Fig.~\ref{fig:ASDs}. These sensitivity curves are subsequently used to generate colored Gaussian noise, into which the simulated CCSN GW signals are injected.

Throughout this work, detector networks are denoted by the concatenation of the corresponding detector labels. For example, HLVKA denotes the five-detector 2G detector network consisting of LIGO Hanford (H), LIGO Livingston (L), Virgo (V), KAGRA (K), and LIGO Aundha (A). Similarly, detector networks containing the 3G detectors are denoted using the corresponding ET and CE labels.

\section{Dataset and methodology}
\label{sec:methods}

\subsection{Data}
\label{sec:data}
The GW signals used in this work are taken from work \citep{Abylkairov26}\footnote{The data are available at https://zenodo.org/records/23302235}. The waveforms were computed with the general relativistic {\tt CoCoNuT} code \citep{Dimmelmeier02a, dimmelmeier:05MdM}. The simulations are axisymmetric, which is adequate during collapse, bounce, and the early post-bounce phase \citep{ott:07cqg}. Deleptonization during collapse is modeled with the density dependent electron fraction prescription $Y_e(\rho)$ \citep{Liebendoerfer05Simple}, and a leakage/heating scheme is used after bounce \citep{Ott13General}.

We consider three solar-metallicity progenitor models with zero-age main sequence (ZAMS) masses of $15$ \citep{Heger05Presupernova}, $27$ \citep{whw:02}, and $40\,M_\odot$ \citep{woosley:07}. These progenitors differ substantially in their core structure: more massive progenitors develop more massive, higher-entropy iron cores, so the set spans a diverse range of pre-collapse conditions \citep{Akhmetali26Toward}.

The progenitors are set into rotation using the cylindrical differential rotation law \citep{abdikamalov:14, richers:17}
\begin{equation}
    \Omega(\varpi) = \Omega_0 \left[ 1 + \left( \frac{\varpi}{A} \right)^2 \right]^{-1},
    \label{eq:rotation_law}
\end{equation}
where $\varpi$ is the cylindrical radius, $\Omega_0$ is the central angular velocity, and $A$ controls the degree of differential rotation. We quantify the impact of rotation by the ratio $T/|W|$, where $T$ is the rotational kinetic energy and $W$ is the gravitational binding energy, both evaluated at core bounce. We consider models with $0.02 \leq T/|W| \leq 0.18$. The lower limit is motivated by the weak GW signal and relatively large numerical noise of slowly rotating models, while the upper limit excludes extremely rapidly rotating models, in which centrifugal support prevents the core from reaching the high densities where differences between EOSs are most pronounced.

For each progenitor, we use the 10 EOSs that yield the most distinguishable GW signals according to the ML-based classification study of \cite{abylkairov2025assessing}: \texttt{LS180}, \texttt{LS220}, \texttt{LS375} \citep{lseos:91}; \texttt{HSTM1}, \texttt{HSDD2}, \texttt{HSIUF} \citep{hempel:10, hempel:12}; \texttt{SFHo}, \citep{steiner:13b}; \texttt{GShenNL3} \citep{gshennl3};  \texttt{GShenFSU2.1} \citep{gshen:11b}; and \texttt{HShen} \cite{hsheneos1,hsheneos2,hshenheos3}. This set covers a broad range of nuclear matter properties and theoretical approaches. After applying the EOS selection, the dataset contains $\mathrm{3790}$ waveforms in total. All waveforms are sampled at $\mathrm{4096}$~Hz.

We restrict the analysis to the interval from $-2$ to $6$~ms relative to the time of bounce. There is little GW signal earlier than $\sim 2$~ms before bounce, while after $\sim 6$~ms the signal begins to include contributions from prompt convection, which is not accurately captured in our axisymmetric simulations \citep{Dolence13Dimensional, Mueller15dynamics}. Hence, this window isolates the bounce and early post-bounce ring-down signal.

\subsection{Signal injection and data preprocessing}
\label{sec:injection}

We generate simulated detector noise using the design sensitivity curves described in Sec.~\ref{sec:detectors} with the \texttt{PyCBC} library~\citep{pycbc}, at a sampling rate of 4096~Hz. We assume the noise to be stationary and Gaussian, with its statistical properties fully specified by the corresponding detector power spectral density (PSD).

We consider a detector network of $N$ detectors, indexed by $k=1,\ldots,N$. For a GW source at sky location $(\theta,\phi)$ with polarization angle $\psi$, the signal measured by detector $k$ is given by
\begin{equation}
\begin{split}
h_k(t)=F^k_{+}(\mathbf{r}_k,\theta,\phi,\psi)\, h_{+}(t-\delta t_k) \\
      +F^k_{\times}(\mathbf{r}_k,\theta,\phi,\psi)\, h_{\times}(t-\delta t_k),
\end{split}
\label{eq:2}
\end{equation}
where $F^k_{+}$ and $F^k_{\times}$ are the antenna pattern functions for the plus and cross polarizations, respectively, $\mathbf{r}_k$ denotes the position of detector $k$ relative to the Earth's center, and $\delta t_k$
is the GW arrival-time delay relative to the Earth's center. The arrival-time delay of the signal at detector $k$ relative to the Earth's center is
\begin{equation}
\delta t_k = \frac{\mathbf{r}_k \cdot \hat{\mathbf{n}}}{c},
\end{equation}
where $\hat{\mathbf{n}}$ is the GW propagation direction and $c$ is the
speed of light. For the axisymmetric 2D CCSN waveforms considered here, $h_{\times}=0$, and hence only the plus-polarization term contributes. The resulting detector data stream is modeled as the sum of the GW signal and detector noise,
\begin{equation}
d_k(t)=n_k(t)+h_k(t),
\end{equation}
where $n_k(t)$ denotes the noise realization in detector $k$.

To account for the dependence of the observed GW amplitude on the source orientation, we consider randomly oriented CCSN sources. For the axisymmetric CCSN simulations considered here, the GW amplitude scales approximately as $\sin^2\alpha$, where $\alpha$ is the angle between the stellar rotation axis and the observer's line of sight~\citep{ott12correlated}. This orientation dependence is incorporated when injecting the simulated waveforms into the detector data.

The CCSN waveforms are injected directly in the time domain. Prior to injection, each waveform is tapered with a Tukey window~\citep{tukey1967intoroduction} with parameter $\alpha=0.1$ to reduce discontinuities at the waveform boundaries and mitigate spectral leakage. The resulting detector strain is then whitened using the corresponding detector PSD and bandpass filtered over the frequency range 20--2000~Hz.

The data are segmented into 70-ms windows, while the CCSN waveforms have a duration of approximately 8 ms. To account for uncertainty in the core-bounce time, we introduce a random offset of $\Delta t_{\rm b}=20$~ms when injecting each signal.

Finally, each 70-ms strain segment is transformed into the frequency domain using a Fourier transform, following our previous study, which demonstrated that the frequency-domain representation provides better performance than the corresponding time-domain representation for EOS classification under bounce-time uncertainty~\citep{Akhmetali26Toward}. The resulting frequency-domain representations are used as inputs to the convolutional neural network (CNN) described in Sec.~\ref{sec:CNN}.

\subsection{Deep learning}
\label{sec:CNN}

Deep learning (DL) has become a powerful approach for extracting complex nonlinear relationships from high-dimensional data~\citep{Lecun15deep}. Among DL methods, convolutional neural networks (CNNs) are particularly well suited to structured signals, as they can automatically learn hierarchical feature representations through convolutional operations~\citep{Gu18CNN, Akhmetali24, Ussipov24}. In this work, we employ a one-dimensional CNN to classify the neutron-star EOS directly from multi-detector GW signals. The network receives frequency-domain strain features, represented as a $144 \times N$ input tensor corresponding to 144 frequency bins across $N$ detector channels, allowing the model to exploit coherent cross-detector spectral features; the full architecture is described in Appendix~\ref{sec:appendix}.

We train the neural network with the Adam optimizer~\citep{kingma17adam} with an initial learning rate of $10^{-3}$ and a batch size of 32. The neural network parameters are optimized by minimizing the sparse categorical cross-entropy loss,
\begin{equation}
\mathcal{L} =
-\frac{1}{M}\sum_{i=1}^{N}
\log p_{i,y_i},
\end{equation}
where $M$ is the number of samples used to compute the loss, $p_{i,y_i}$ is the probability assigned by the neural network to the true EOS class $y_i$ for the $i$th sample, and $y_i$ denotes the corresponding ground-truth EOS class. Training is performed for a maximum of 400 epochs. To facilitate convergence, the learning rate is reduced by a factor of two if the validation loss does not improve for 20 consecutive epochs, subject to a minimum learning rate of $10^{-6}$. We additionally employ early stopping with a patience of 40 epochs and restore the model weights corresponding to the lowest validation loss. These measures help prevent overfitting while allowing the neural network to adequately optimize its parameters. The dataset of 3790 signals is randomly split into 80\% training and 20\% validation subsets. Before training, each detector channel is standardized independently by subtracting its mean and dividing by its standard deviation. The scaling parameters are determined from the training set using {\tt StandardScaler} from {\tt scikit-learn}~\citep{pedregosa2011scikit} and subsequently applied to both the training and test data. 

We quantify the EOS classification performance using the accuracy metric,
\begin{equation}
\text{Accuracy} = \frac{\text{Number of Correct Predictions}}{\text{Total Number of Predictions}} \,.
\end{equation}
We adopt an accuracy threshold of 0.7 as the criterion for sufficiently reliable EOS classification and use this threshold to define the effective distance reach of the detector network.
To reduce the dependence of the results on a particular train--test partition, we repeat the complete training and evaluation procedure 20 times. For each iteration, a new stratified random split is generated and the CNN is retrained from scratch. The reported classification accuracy is the mean over the 20 independent runs, while the associated uncertainty is given by their standard deviation. This procedure provides a more robust estimate of the model performance and reduces the influence of statistical fluctuations associated with any individual data partition or neural network initialization. Further details on the network architecture, training behaviour, and classification performance are provided in Appendix~\ref{sec:appendix}.

\section{Results}
\label{sec:results}

To evaluate the performance of the EOS classification framework, we use the CCSN waveforms described in Sec.~\ref{sec:data} across different detector network configurations and source locations. In all experiments, we perform 10-class EOS classification. The classification accuracy depends on several properties of the detector--source configuration, in particular the source distance, the GPS arrival time, and the sky location. To investigate the influence of these factors independently, we vary one parameter at a time while keeping the other two fixed. We first examine the dependence of the classification accuracy on the source distance, followed by its dependence on the signal arrival time and sky location. Finally, we extend the analysis to third-generation detector networks to assess the performance expected with improved detector sensitivity.

\subsection{Dependence on distance}
\label{sec:Distance}

We first assess the dependence of the EOS classification performance on the distance to the source for different configurations of the 2G detector network. As the distance increases, the SNR decreases, making the classification increasingly challenging. For each detector network configuration, we fix the source position at RA$=18.34$ h and Dec$=-16.18^\circ$ within the Sagittarius constellation, consistent with the expectation that Galactic CCSNe predominantly occur near the Galactic plane. We then vary the source distance to assess the detection reach of each network configuration.

Because the antenna response of each interferometer depends on the GW propagation direction relative to its arm orientation, Earth’s rotation modulates the observed signal amplitude and may consequently influence classification performance. Following the work of~\citet{Bruel23Inference}, we consider two cases for each detector network configuration, corresponding to favourable and unfavourable antenna responses. These cases are determined by the equivalent antenna pattern of a detector network consisting of $N$ detectors, defined as
\begin{equation}
F_{\rm eq} =
\sqrt{\frac{1}{N}
\sum_{k\in\mathrm{network}}
\left[
F_{+,k}^2(\vec{\Omega},t_0)
+
F_{\times,k}^2(\vec{\Omega},t_0)
\right]},
\end{equation}
where $\vec{\Omega}$ denotes the source sky position and $t_0$ is the GW arrival time at the centre of the Earth. As mentioned before, in our axisymmetric simulations $F_{\times}=0$, and therefore the equivalent antenna pattern reduces to
\begin{equation}
F_{\rm eq} =
\sqrt{\frac{1}{N}
\sum_{k\in\mathrm{network}}
F_{+,k}^2(\vec{\Omega},t_0)}.
\end{equation}
The favourable case corresponds to an arrival time that maximizes $F_{\rm eq}$ for the fixed sky position over a 24-hour period, while the unfavourable case corresponds to the arrival time that minimizes $F_{\rm eq}$ over the same period. The corresponding arrival times and values of $F_{\rm eq}$ for the different detector network configurations are given in Table~\ref{tab:antenna_cases}.

We note that this definition of favourable and unfavourable cases accounts only for the antenna patterns of the interferometers and does not consider their differing sensitivities. Thus, the favourable case does not necessarily correspond to the maximum achievable network SNR, but instead provides a reference configuration for comparison with the less favourable case.

\begin{table}[htbp]
\centering
\caption{Favourable and unfavourable arrival times and corresponding equivalent antenna pattern values, $F_{\rm eq}$, for the different 2G detector-network configurations. The favourable and unfavourable cases are determined by maximizing and minimizing $F_{\rm eq}$, respectively, over a 24-hour period for the fixed source sky position starting from an arbitrary time $t_{\rm GPS}=1325048418$.}
\label{tab:antenna_cases}
\begin{tabular}{ccccc}
\hline
Network & Favourable & F$_{\rm eq}$ & Unfavourable & F$_{\rm eq}$ \\
\hline
HL    & 1325130947 & 0.84 & 1325059908 & 0.09 \\
HLV   & 1325130202 & 0.69 & 1325059541 & 0.09 \\
HLVK  & 1325129912 & 0.60 & 1325058582 & 0.16 \\
HLVKA & 1325130858 & 0.55 & 1325103368 & 0.17 \\
\hline
\end{tabular}
\end{table}

\begin{figure*}[t]
\centering
\includegraphics[width=1\linewidth]{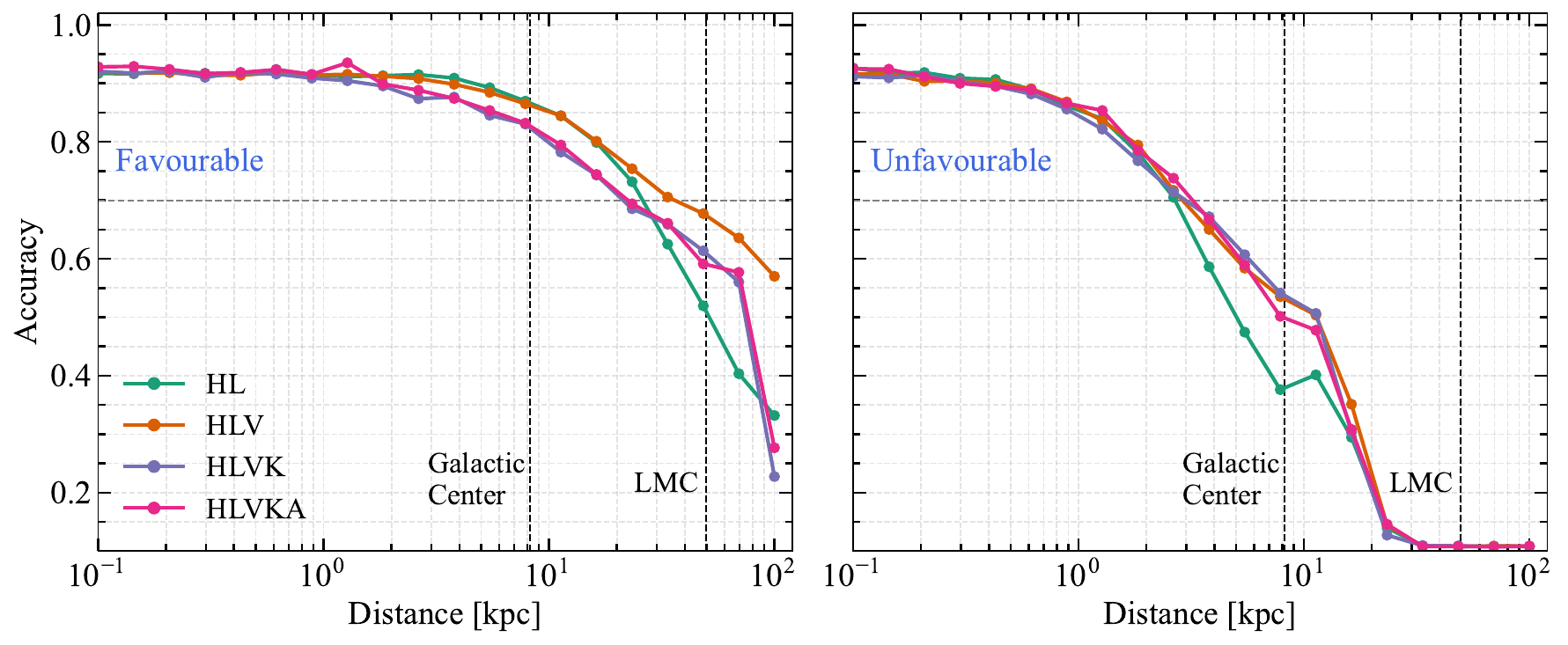}
    \caption{Classification accuracy as a function of the source distance for different 2G detector network configurations. The left and right panels correspond to favourable and unfavourable GW arrival times, respectively (see Table~\ref{tab:antenna_cases} and the main text for details). Each curve corresponds to a different detector network configuration: HL, HLV, HLVK, HLVA, and HLVKA, as indicated in the legend. The gray horizontal dashed line at 0.7 indicates the threshold adopted for reasonably accurate classification performance. The vertical dashed lines mark the distances to the Galactic Center (8.2 kpc) and the Large Magellanic Cloud (LMC; 49.8 kpc).}
     \label{fig:distance}
\end{figure*}

Figure~\ref{fig:distance} shows the evolution of the classification accuracy as a function of the source distance. We observe a substantial difference between the favourable and unfavourable cases. For favourable arrival times, the 2G detector networks achieve a distance reach of $\sim 20$ kpc, whereas for unfavourable arrival times, the reach is reduced to only $\sim 3$ kpc, independent of number of detectors in the detector network. We note that, in the favourable case, adding Virgo to the HL network substantially improves the distance reach, extending it by about 10 kpc. In contrast, the HLVK and HLVKA networks do not improve the reach relative to HL at the adopted accuracy threshold. This behaviour may be related to the $F_{\rm eq}$ (see Table~\ref{tab:antenna_cases}), which is substantially lower for the selected HLVK and HLVKA configurations than for HL, indicating a weaker effective network response for the corresponding source-detector geometries.

Overall, the current detector network enables accurate EOS classification of rotating Galactic CCSNe when the signal arrival time is favourable.

\subsection{Dependence on arrival time}
\label{sec:Time}

We now examine the effect of the Earth's rotation on the accuracy of EOS classification. We fix the source at the Galactic Center (RA$=18.34$ h, Dec$=-16.18^\circ$, and $d=8.2$ kpc). To account for Earth's rotation, we vary the GW arrival time at Earth over a 24-hour period, starting from an arbitrary reference time $t_{\rm GPS}=1325048418$.

\begin{figure}[htbp]
\centering
\includegraphics[width=1\linewidth]{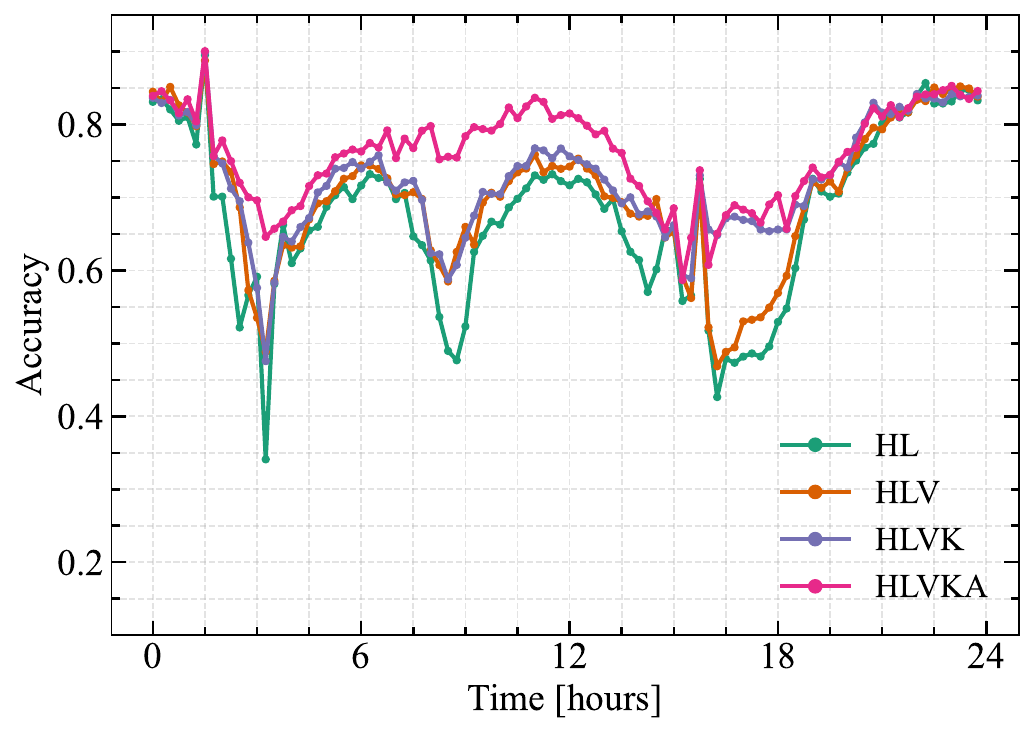}
    \caption{Classification accuracy as a function of the GW arrival time over a 24-hour period for different 2G detector-network configurations. Each curve corresponds to a different detector network configuration: HL, HLV, HLVK, HLVA, and HLVKA, as indicated in the legend.}
     \label{fig:time}
\end{figure}

Figure~\ref{fig:time} shows the evolution of the classification accuracy as a function of the arrival time. Starting with the HL detector network, consisting of the two nearly co-aligned LIGO Hanford and LIGO Livingston detectors, adding Virgo improves the classification performance between $t=7.5$ h and $t=15$ h. The addition of KAGRA further improves the accuracy between $t=16$ h and $t=18$ h, when the other detectors have a substantially reduced response to the signal. Finally, the addition of LIGO Aundha, which has a sensitivity comparable to that of the LIGO detectors in the HL detector network (see Fig.~\ref{fig:ASDs}), improves the classification performance across most of the 24-hour period. In particular, LIGO Aundha provides additional sensitivity around $t=3$ h, when the other detectors have a weak response to the signal, and contributes to a systematic improvement in accuracy between $t=6$ h and $t=15$ h. We note that \citet{Bruel23Inference} report two pronounced dips in the coverage over a 24-hour period (their Fig.~8), whereas three dips are visible in our results. This difference arises from the axisymmetric waveform assumption adopted in our analysis, for which the cross-polarisation component is omitted from Eq.~(6), resulting in a different modulation of the detector response with the source arrival time.

To further quantify the improvement in classification performance over the 24-hour period, we calculate the normalized area under curve (AUC) for Fig.~\ref{fig:time}, which provides a measure of the average classification accuracy over a full day. The resulting normalized AUC values are $0.67$, $0.70$, $0.72$, and $0.76$ for the HL, HLV, HLVK, and HLVKA detector networks, respectively. These results show a systematic improvement in the overall classification performance as additional detectors are incorporated into the detector network. In particular, the HLVKA detector network achieves the highest day-averaged accuracy, demonstrating that the additional detector coverage helps mitigate the loss of sensitivity caused by the Earth's rotation.

Overall, these results demonstrate that the Earth's rotation has a significant impact on the EOS classification performance by periodically changing the antenna response of the detector network. While the classification accuracy can vary substantially over a 24-hour period for a given detector network, adding detectors improves the overall performance and provides more uniform coverage in time. 

\subsection{Dependence on sky location}
\label{sec:Sky}

We next investigate the dependence of EOS classification accuracy on the source sky location and assess how the addition of detectors affects the uniformity of classification performance across the sky. Because the detector response is determined by the antenna pattern, which varies with the source position relative to each detector (see Eq.~(\ref{eq:2})), the observed signal strength and morphology can differ substantially across the celestial sphere. Consequently, classification performance is expected to exhibit a corresponding dependence on sky location. Expanding the detector network introduces additional, differently oriented and geographically separated detectors, which is expected to provide complementary responses to the GW signal and thereby improve the overall sky coverage and robustness of the classification.

\begin{figure*}[ht]
\centering
\includegraphics[width=1\linewidth]{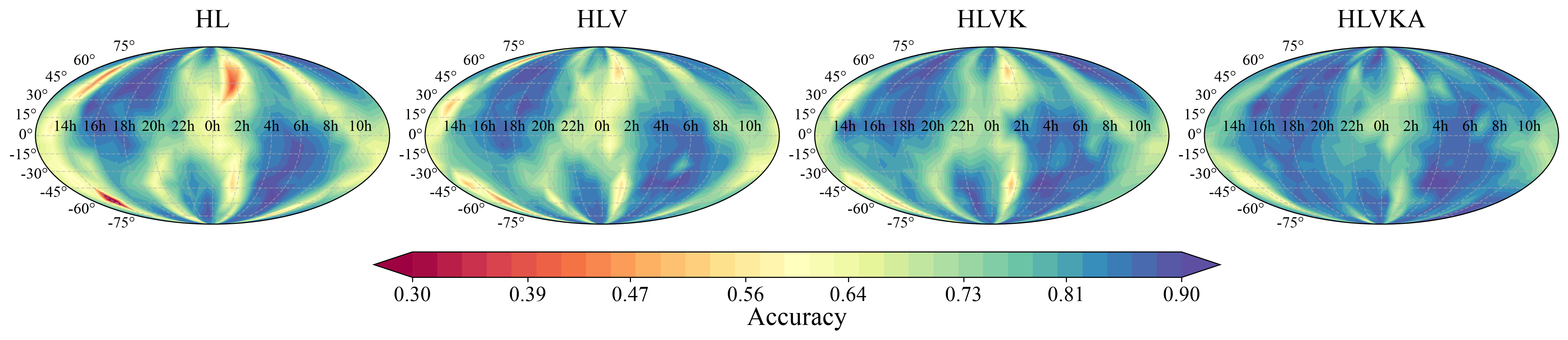}
    \caption{Sky maps of the accuracy score for different detector network configurations of 2G detectors. Titles correspond to the detector networks, while colors indicate the classification performance, with warmer colors representing lower accuracy values and cooler colors representing higher accuracy values.} 
     \label{fig:skymap}
\end{figure*}

For this experiment, we fix the source distance to the Galactic Center (8.2 kpc) and the arrival time to a favourable value (see Table~\ref{tab:antenna_cases}), while varying the source sky location isotropically. Figure~\ref{fig:skymap} shows the distribution of classification accuracy across the sampled sky positions. Because the detector response varies with sky location, the classification performance exhibits localized regions of reduced accuracy associated with minima in the detector network antenna response. As additional interferometers are included, these low-performance regions become progressively smaller, resulting in more uniform classification performance across the sky.

\begin{figure}[htbp]
\centering
\includegraphics[width=1\linewidth]{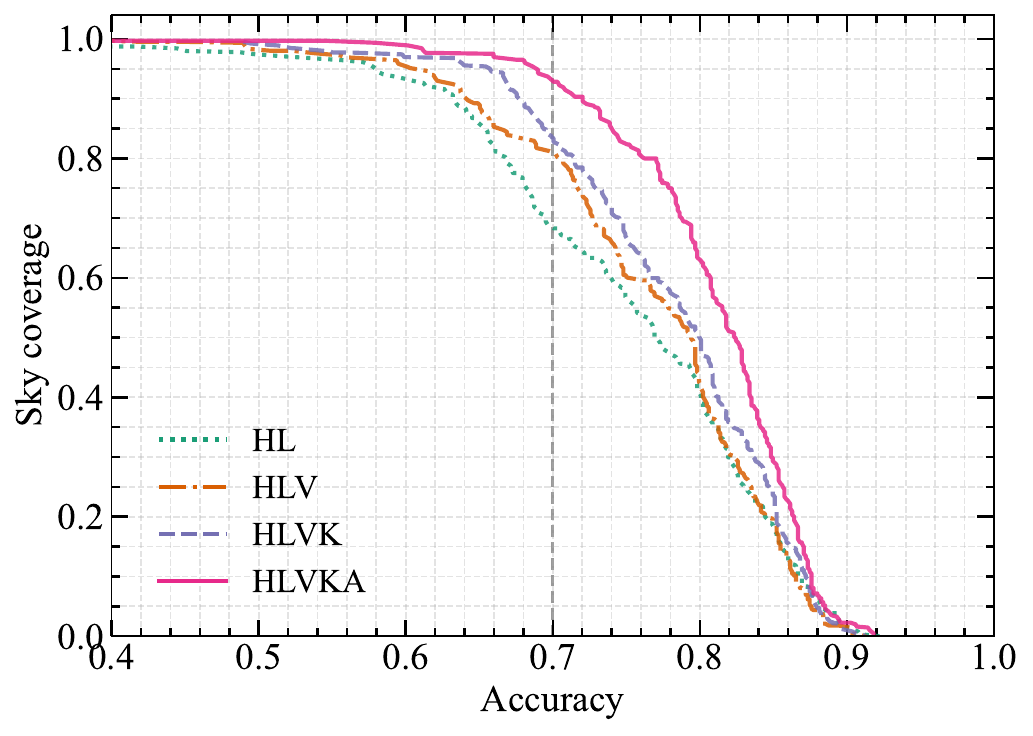}
    \caption{Sky coverage as a function of the classification accuracy threshold for different 2G detector network configurations. Different colors denote the HL, HLV, HLVK, and HLVKA detector networks. The gray vertical dashed line at an accuracy of 0.7 marks the adopted threshold for accurate classification and provides a reference for comparing the sky coverage achieved by each detector network.} 
     \label{fig:coverage}
\end{figure}

To quantify this improvement, Fig.~\ref{fig:coverage} presents the fraction of the sky over which the classification accuracy exceeds a given threshold. The results show a systematic increase in sky coverage with the addition of detectors. At an accuracy threshold of 0.7, the sky coverage increases from 0.69 for the HL detector network to 0.81, 0.84, and 0.93 for the HLV, HLVK, and HLVKA detector networks, respectively. The inclusion of Virgo produces the largest single improvement, increasing the sky coverage by $\sim 12 \%$. 

Overall, expanding the detector network reduces the directional dependence of the detector response, thereby yielding more uniform classification performance across the sky.

\subsection{Third-generation detectors}
\label{sec:3G}

Finally, we investigate the distance reach of the classification framework with 3G detectors. Owing to their substantially improved sensitivity compared with current 2G observatories, 3G detectors are expected to extend the distance over which CCSN signals can be detected and reliably classified.

\begin{figure*}[ht]
\centering
\includegraphics[width=1\linewidth]{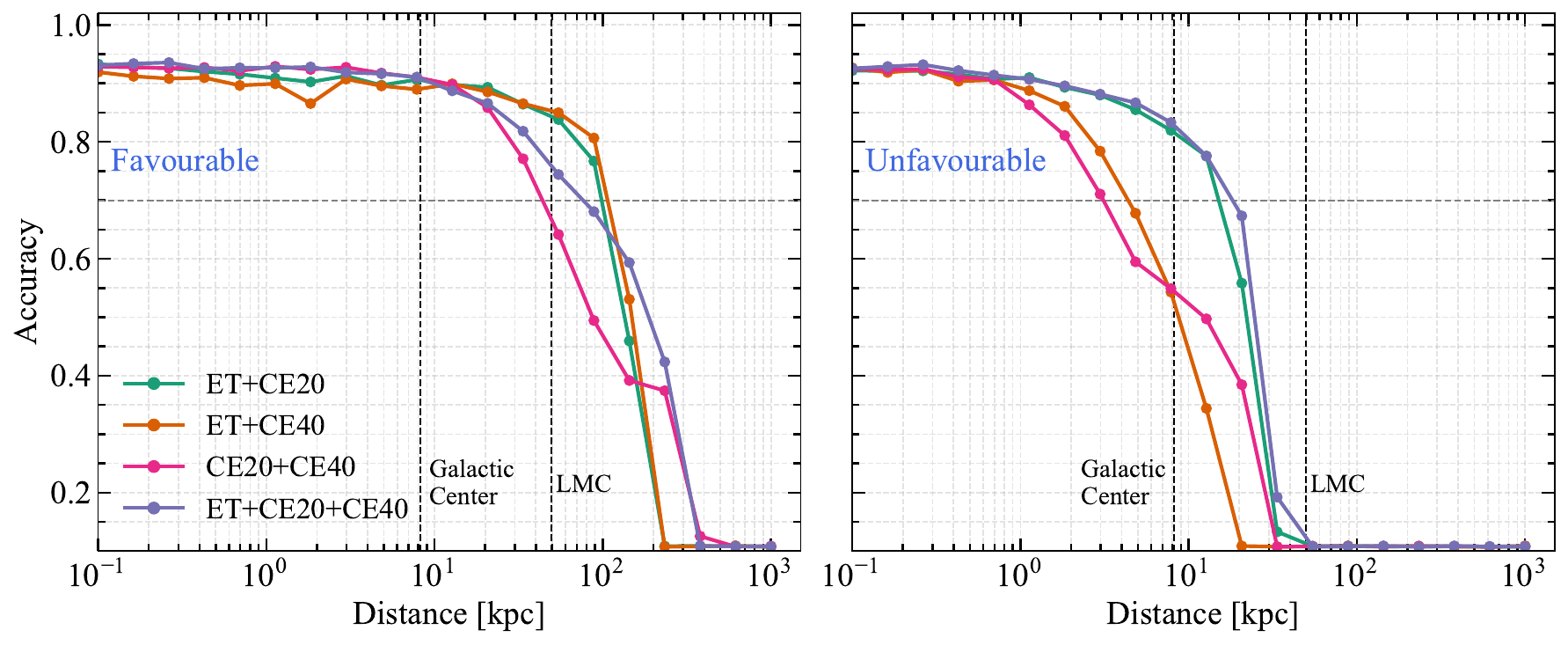}
    \caption{Classification accuracy as a function of the source distance for different 3G detector network configurations. The left and right panels correspond to favourable and unfavourable GW arrival times, respectively. Each curve corresponds to a different detector network configuration, as indicated in the legend. The gray horizontal dashed line at 0.7 indicates the threshold adopted for reasonably accurate classification performance. The vertical dashed lines mark the distances to the Galactic Center (8.2 kpc) and the LMC (49.8 kpc).}
     \label{fig:3G}
\end{figure*}

We repeat the distance-dependent analysis described in Sec.~\ref{sec:Distance} using detector networks composed of the 3G observatories ET, CE20, and CE40. As in Sec.~\ref{sec:Distance}, we fix the source location to the Galactic Center and consider both favourable and unfavourable signal arrival times. The corresponding GPS times and detector network antenna response factors are listed in Table~\ref{tab:3G}. We consider the detector networks ET+CE20, ET+CE40, CE20+CE40, and the full ET+CE20+CE40 configuration. Figure~\ref{fig:3G} shows the classification performance as a function of source distance.

As for the 2G detector networks, we find a substantial dependence of the distance reach on the signal arrival time. For the favourable cases, the CE20+CE40 detector network provides the shortest classification reach, extending to approximately the distance of the LMC, whereas the ET+CE40 detector network achieves the largest reach of $\sim$100~kpc. For the unfavourable cases, the corresponding reaches are substantially reduced: the CE20+CE40 detector network reaches $\sim 3$~kpc, while the full ET+CE20+CE40 detector network extends the classification reach to $\sim 20$~kpc.

Overall, the 3G detector networks provide a substantial improvement in the distance reach compared with the 2G configurations, particularly for favourable signal arrival times. However, the strong difference between favourable and unfavourable cases demonstrates that the increased detector sensitivity does not eliminate the dependence of CCSN classification performance on the detector network antenna response. Adding multiple 3G detectors with different orientations and locations is therefore expected to mitigate this directional dependence and improve the robustness of the classification framework.

\begin{table}[ht]
\centering
\caption{As in Table~\ref{tab:antenna_cases}, but for detector networks consisting of 3G detectors.}
\label{tab:3G}
\begin{tabular}{ccccc}
\hline
Network & Favourable & F$_{\rm eq}$ & Unfavourable & F$_{\rm eq}$ \\
\hline
ET+CE20     & 1325111125 & 0.52 & 1325081840 & 0.07 \\
ET+CE40     & 1325110582 & 0.51 & 1325083158 & 0.05 \\
CE20+CE40   & 1325050485 & 0.77 & 1325110252 & 0.05 \\
ET+CE20+CE40 & 1325049352 & 0.49 & 1325082259 & 0.07 \\
\hline
\end{tabular}
\end{table}

\section{Conclusion}
\label{sec:conclusion}

In this work, we have presented a DL framework for classifying the nuclear EOS of rotating CCSNe from their GW signals. Building upon our previous single-detector EOS classification study~\citep{Akhmetali26Toward}, we have extended the framework to arbitrary detector networks of GW interferometers and investigated how the classification performance depends on the source distance, GW arrival time, and sky location. Using CNNs, we performed 10-class EOS classification directly from noisy detector data and assessed the performance for different configurations of second- and third-generation detectors.

Our results show that the classification performance is strongly influenced by the detector--source geometry. For current 2G detector networks, accurate EOS classification of Galactic CCSNe is achievable when the signal arrives at a favourable time. However, the distance reach remains strongly dependent on the detector network antenna response, and adding detectors does not necessarily extend the reach, as the resulting performance depends on both the sensitivity and orientation of the individual interferometers (cf.~Sec.~\ref{sec:Distance}).

The effect of the detector--source geometry is particularly evident when considering the Earth's rotation. For a fixed source location, the classification accuracy can vary substantially over a 24-hour period as the antenna response of the detector network changes. Nevertheless, expanding the detector network improves the overall time-averaged performance. The normalized AUC increases from $0.67$ for the HL detector network to $0.70$, $0.72$, and $0.76$ for the HLV, HLVK, and HLVKA detector networks, respectively. The inclusion of geographically separated and differently oriented detectors provides complementary responses to the GW signal, reducing periods of poor detector network sensitivity and yielding more uniform classification performance throughout the day (cf.~Sec.~\ref{sec:Time}).

The dependence on sky location shows a similar behaviour. For a fixed source distance and arrival time, the classification accuracy exhibits localized regions of reduced performance associated with minima in the detector network antenna response. Adding detectors progressively reduces these low-performance regions and produces more uniform classification performance across the sky. At an accuracy threshold of $0.7$, the fraction of the sky satisfying this criterion increases from $0.69$ for the HL detector network to $0.81$, $0.84$, and $0.93$ for the HLV, HLVK, and HLVKA detector networks, respectively. The inclusion of Virgo provides the largest single improvement, increasing the sky coverage by 0.12 points. These results demonstrate that the principal benefit of expanding the 2G detector network is not necessarily a substantial increase in distance reach, but rather a reduction in the directional dependence of the classification performance (cf.~Sec.~\ref{sec:Sky}).

Finally, we have assessed the potential of 3G detector networks consisting of ET and CE observatories. We find that the improved sensitivity of 3G detectors extends the distance over which accurate EOS classification can be achieved. For favourable arrival times, the classification reach extends to $\sim 100$~kpc for the ET+CE40 detector network. For unfavourable arrival times, the reach is considerably reduced to $\sim 20$~kpc for the full ET+CE20+CE40 detector network. Thus, although the increased sensitivity of 3G detectors substantially extends the classification reach, it does not eliminate the dependence on the detector antenna response. The combination of multiple 3G observatories is therefore expected to be important for obtaining robust EOS classification over a broad range of source locations and arrival times (cf.~Sec.~\ref{sec:3G}).

Overall, our results demonstrate that DL can provide an effective approach for rapid EOS classification of rotating CCSNe directly from noisy GW detector data. The results further show that the performance of such a framework is determined not only by the intrinsic sensitivity of the detectors, but also by the detector network geometry and the relative orientation of the source and detector network. Increasing the number of detectors can therefore provide two complementary benefits: improved sensitivity and, importantly, more uniform coverage in time and across the sky. This is particularly relevant for CCSNe, for which the explosion time and sky location are generally unknown in advance, making it impossible to select a detector configuration with a favourable antenna response beforehand.

Several limitations remain to be addressed. Our analysis is restricted to rotating CCSN models and considers a finite set of 10 EOS classes, while the diversity of realistic progenitors and explosion dynamics is substantially broader. In addition, the analysis focuses on the GW signal around core bounce and the early post-bounce phase and does not consider the full duration and complexity of CCSN GW emission. The detector noise is also assumed to be stationary and Gaussian, and therefore does not capture the non-stationary and non-Gaussian noise transients present in real detector data. An additional limitation arises from the axisymmetric nature of the CCSN simulations considered here. In these models, only the plus polarization is present, and therefore the detector response is determined solely by the corresponding antenna pattern. Fully three-dimensional CCSN simulations can produce both GW polarizations, which may modify the detector-network response and, consequently, the time- and sky-dependent classification performance. The magnitude of this effect requires investigation with fully three-dimensional CCSN models and is left for future work.

Likewise, future work should address these limitations by incorporating a broader range of progenitor models, fully three-dimensional simulations, more realistic detector noise, and additional features of the CCSN signal. Extending the framework to realistic, continuously operating detector networks and combining EOS classification with simultaneous source localization and physical-parameter inference will provide a further step toward the real-time interpretation of CCSN GW observations.

\begin{acknowledgments}
This research was funded by the Science Committee of the Ministry of Science and Higher Education of the Republic of Kazakhstan (Grant No. AP26103591). EA is partially supported by the Nazarbayev University Faculty Development Competitive Research Grant Program (no. 040225FD4713). 
JAF is supported by the Spanish Agencia Estatal de Investigación (grant PID2024-159689NB-C21) funded by MICIU/AEI/10.13039/501100011033 and by FEDER / EU, and by the Generalitat Valenciana (Prometeo Excellence Programme grant CIPROM/2022/49).
In preparing this manuscript the authors used ChatGPT for language editing and improving grammar. The analysis, the results and the conclusions are the authors’ own, and the authors take full responsibility for the content of this work.
\end{acknowledgments}

\appendix

\section{CNN model}
\label{sec:appendix}

We provide here additional details on the CNN architecture, its training behaviour, and its performance in the EOS classification task.

\subsection{CNN architecture}

Table~\ref{tab:cnn} summarizes the CNN architecture adopted in this study. The network consists of three convolutional blocks, each comprising a Conv1D layer followed by a MaxPooling1D layer. The convolutional layers use 32, 64, and 128 filters, respectively, with a kernel size of 3 and ReLU activation. These layers progressively extract features from the frequency-domain GW signals, while the pooling layers reduce the dimensionality of the feature representation. L2 regularization with a coefficient of $10^{-2}$ is applied to the convolutional and fully connected layers to mitigate overfitting.

Following the final convolutional block, the resulting feature maps are flattened into a 2048-dimensional vector and passed to a fully connected layer with 64 neurons and ReLU activation. The final layer contains 10 neurons, corresponding to the EOS classes considered in this study, and uses a softmax activation to produce class probabilities. The class with the highest predicted probability is taken as the classification result. The resulting network contains 162,922 trainable parameters.

\begin{table*}[htbp]
\centering
\caption{Architecture of the CNN model used in this work.}
\label{tab:cnn}
\begin{tabular}{lllcc}
\hline
Layer & Type & Parameters & Output Shape & Activation \\
\hline
0 & Conv1D & 32 filters, kernel size 3 & $(142,32)$ & ReLU \\
1 & MaxPooling1D & Pool size 2 & $(71,32)$ & -- \\
2 & Conv1D & 64 filters, kernel size 3 & $(69,64)$ & ReLU \\
3 & MaxPooling1D & Pool size 2 & $(34,64)$ & -- \\
4 & Conv1D & 128 filters, kernel size 3 & $(32,128)$ & ReLU \\
5 & MaxPooling1D & Pool size 2 & $(16,128)$ & -- \\
6 & Flatten & -- & $(2048)$ & -- \\
7 & Dense & 64 units & $(64)$ & ReLU \\
8 & Dense & 10 units & $(10)$ & Softmax \\
\hline
\end{tabular}
\end{table*}

\subsection{Training behaviour}

\begin{figure*}[t]
\centering
\includegraphics[width=1\linewidth]{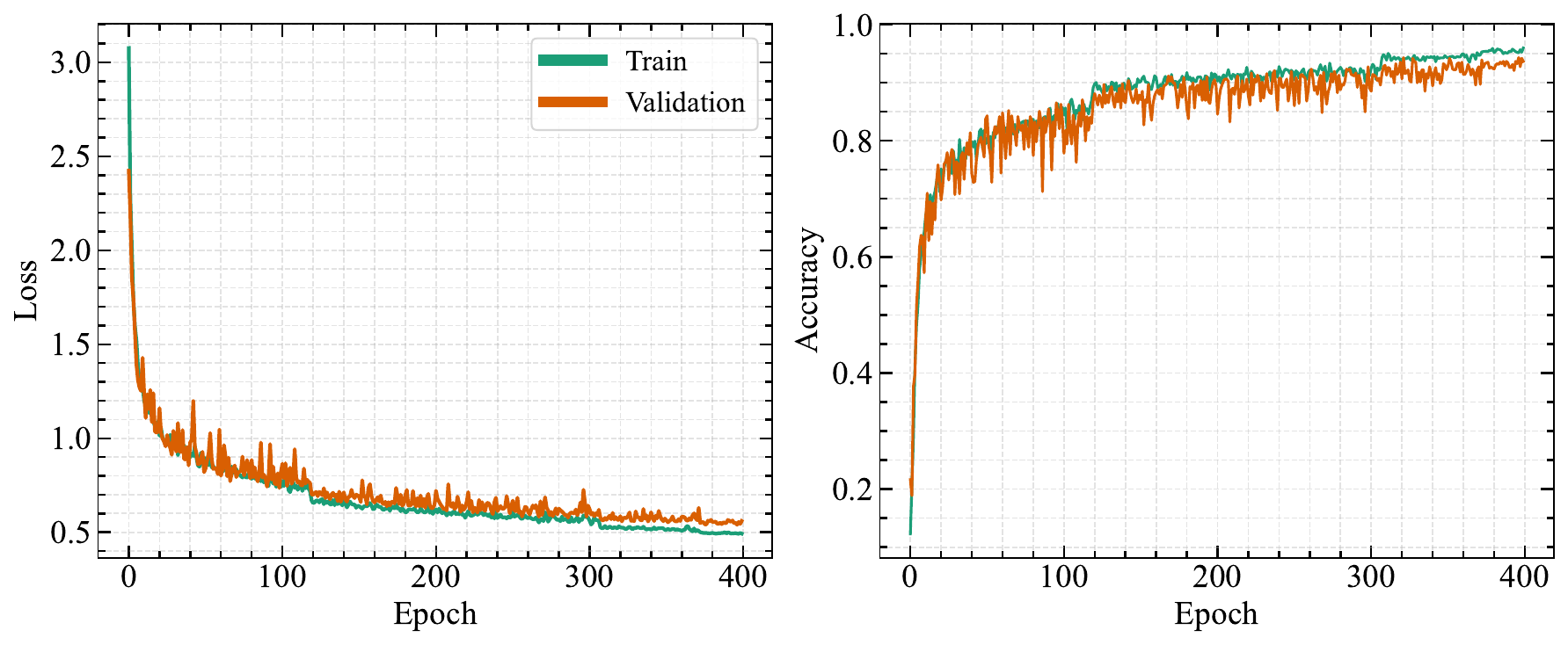}
\caption{Training and validation loss and accuracy as a function of epoch for a representative CNN training run.}
\label{fig:CNN_train}
\end{figure*}

As a representative test, we inject GW signals at a distance of 2~kpc, assuming the Galactic Centre sky position and a favourable detector response for the HL network (see Sec.~\ref{sec:Distance} for details). Figure~\ref{fig:CNN_train} shows the corresponding training and validation loss and accuracy as a function of epoch. The model improves rapidly during the initial stage of training and subsequently converges more gradually as the learning rate is reduced. The training and validation curves follow similar trends, with only a small separation between them.

Some epoch-to-epoch fluctuations are visible, particularly in the validation metrics, as expected from the stochastic nature of mini-batch optimization. However, the validation loss shows no sustained increase, while the validation accuracy remains stable after reaching its plateau. This behaviour indicates stable convergence without pronounced overfitting.

The final model weights are selected according to the minimum validation loss using the early-stopping procedure described in Section~\ref{sec:CNN}. In the representative run shown in Figure~\ref{fig:CNN_train}, early stopping is not triggered before the maximum number of epochs is reached. Overall, the training history demonstrates that the CNN converges reliably while maintaining good performance on the validation data.

\subsection{Classification performance}

\begin{figure}[htbp]
\centering
\includegraphics[width=1\linewidth]{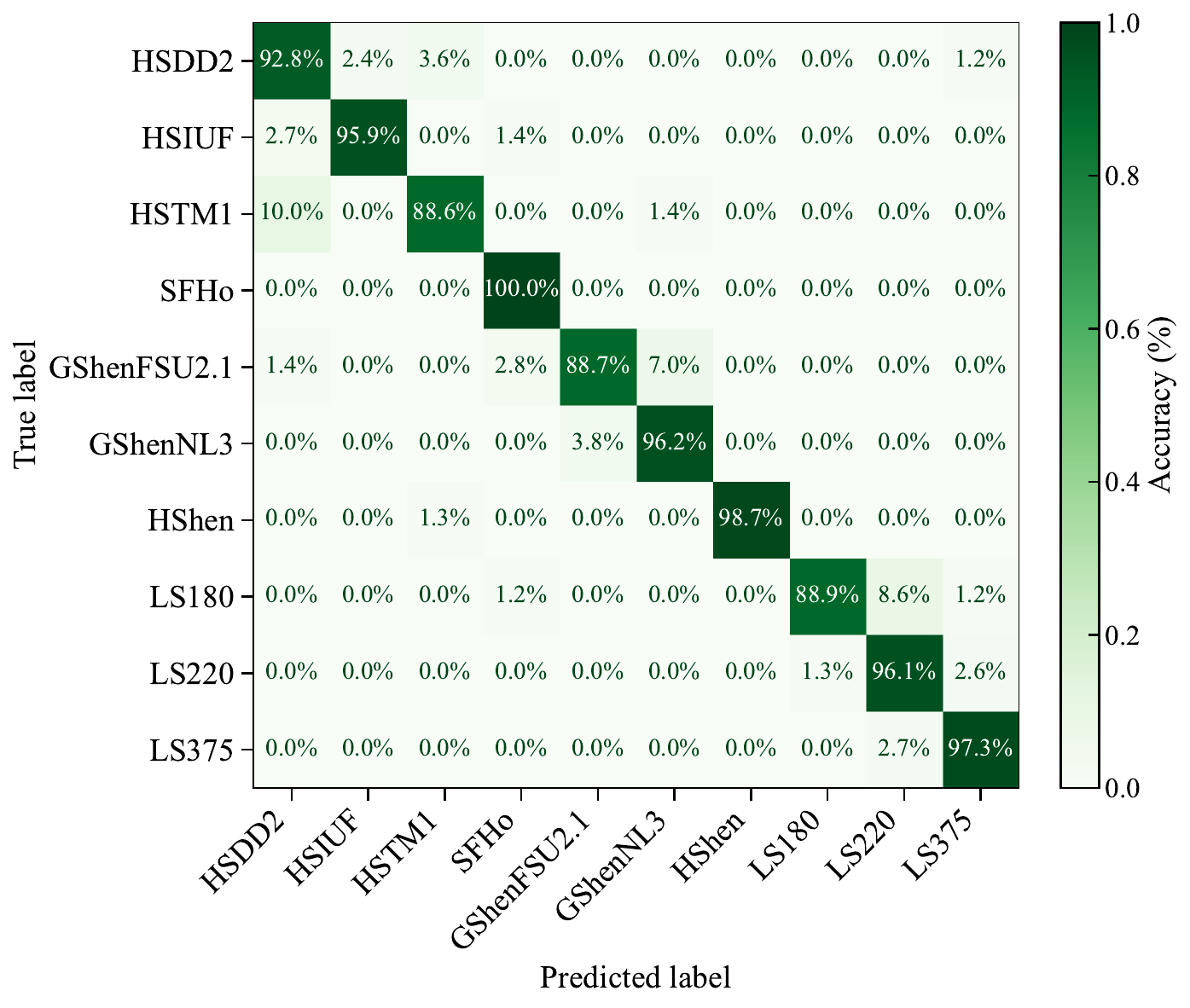}
\caption{Confusion matrix for the 10-EOS classification for the HL network at a source distance of 2~kpc, favourable case and a Galactic Center sky location. Classification accuracy is 94\%.}
\label{fig:CNN_cm}
\end{figure}

The classification performance for the same representative configuration is shown in Figure~\ref{fig:CNN_cm}. For the full 10-EOS classification, the HL network achieves an accuracy of 94\% at a source distance of 2~kpc, assuming a favourable orientation and a Galactic-center sky location. Most predictions lie along the diagonal of the confusion matrix, while the remaining misclassifications are concentrated among a small number of EOS pairs. This indicates that these classes produce similar GW signal morphologies and are consequently more difficult for the CNN to distinguish.

The classification accuracy therefore does not reach 100\% even for this nearby source and favourable detector response. Reducing the relative contribution of detector noise through a larger signal amplitude cannot fully eliminate the ambiguity between intrinsically similar EOS waveforms. The remaining classification errors thus appear to be primarily associated with degeneracies between specific EOS classes rather than with insufficient signal sensitivity.

\begin{figure}[htbp]
\centering
\includegraphics[width=1\linewidth]{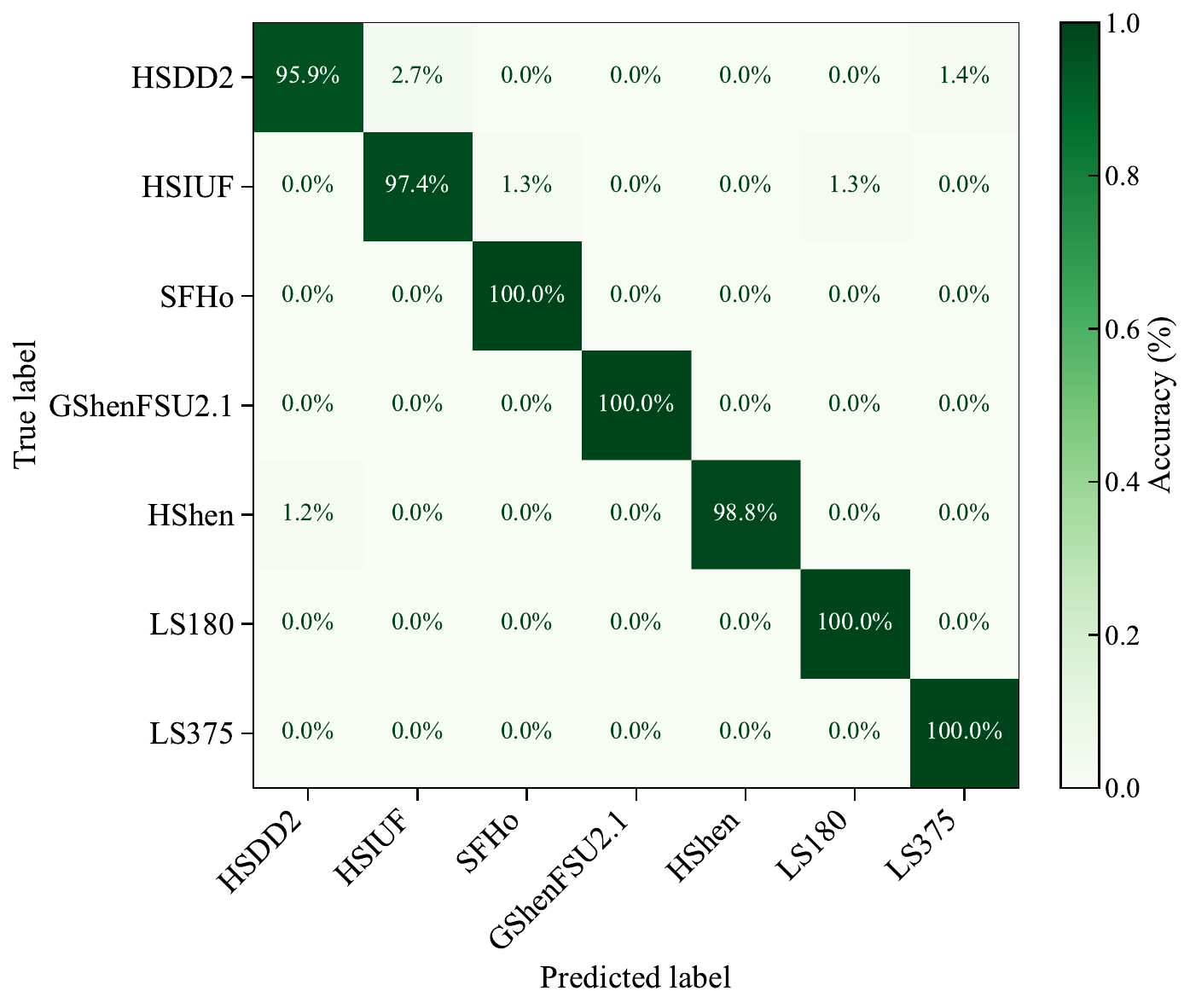}
\caption{Confusion matrix for the reduced 7-EOS classification after excluding HSTM1, GShenNL3, and LS220, which show the strongest mutual confusions in the full classification. Classification accuracy is 99\%.}
\label{fig:CNN_cm_reduced}
\end{figure}

To further examine this effect, we exclude the three EOS classes associated with the strongest confusions, HSTM1, GShenNL3, and LS220, and repeat the classification for the remaining seven EOSs using the same injection configuration. The resulting confusion matrix is shown in Figure~\ref{fig:CNN_cm_reduced}. The classification accuracy increases to 99\%, with substantially fewer off-diagonal predictions. This improvement supports the interpretation that the residual errors in the full 10-EOS classification are mainly driven by similarities between particular EOS classes rather than by a fundamental limitation of the CNN or the HL detector network.

\bibliography{gw_sn}

\end{document}